\documentclass[twocolumn,secnumarabic,amssymb, nobibnotes, aps, pra]{revtex4-2}
\usepackage{lineno}

\usepackage{times}
\usepackage[pdftex]{graphicx}
\usepackage{dcolumn}
\usepackage{subfig}
\usepackage{bm}
\usepackage{amsmath}
\usepackage{indentfirst}
\usepackage{float}
\usepackage[colorlinks]{hyperref}
\usepackage[dvipsnames]{xcolor}
\usepackage{xcolor}
\usepackage[no-test-for-array]{nicematrix}
\usepackage{tikz}
\usepackage{lipsum}
\usepackage{makecell}
\usepackage[normalem]{ulem}
\usepackage[font=small,justification=RaggedRight]{caption}
\usepackage{braket}

\newcommand{\im}{\mathrm{i}}
\newcommand{\rme}{\mathrm{e}}

\begin{document}

\title{Thermometry of Trapped Ions Based on Bichromatic Driving}

\author{Xie-Qian Li}
\thanks{These authors contribute equally to this work.}
\affiliation{Institute for Quantum Science and Technology, National University of Defense Technology, Changsha Hunan 410073, China}
\affiliation{Hunan Key Laboratory of Mechanism and Technology of Quantum Information, Changsha Hunan 410073, China}

\author{Yi Tao}
\thanks{These authors contribute equally to this work.}
\affiliation{Institute for Quantum Science and Technology, National University of Defense Technology, Changsha Hunan 410073, China}
\affiliation{Hunan Key Laboratory of Mechanism and Technology of Quantum Information, Changsha Hunan 410073, China}

\author{Ting Chen}
\affiliation{Institute for Quantum Science and Technology, National University of Defense Technology, Changsha Hunan 410073, China}
\affiliation{Hunan Key Laboratory of Mechanism and Technology of Quantum Information, Changsha Hunan 410073, China}

\author{Wei Wu}
\email{weiwu@nudt.edu.cn}
\affiliation{Institute for Quantum Science and Technology, National University of Defense Technology, Changsha Hunan 410073, China}
\affiliation{Hunan Key Laboratory of Mechanism and Technology of Quantum Information, Changsha Hunan 410073, China}

\author{Yi Xie}
\email{xieyi2015@nudt.edu.cn}
\affiliation{Institute for Quantum Science and Technology, National University of Defense Technology, Changsha Hunan 410073, China}
\affiliation{Hunan Key Laboratory of Mechanism and Technology of Quantum Information, Changsha Hunan 410073, China}

\author{Chun-Wang Wu}
\email{cwwu@nudt.edu.cn}
\affiliation{Institute for Quantum Science and Technology, National University of Defense Technology, Changsha Hunan 410073, China}
\affiliation{Hunan Key Laboratory of Mechanism and Technology of Quantum Information, Changsha Hunan 410073, China}

\author{Ping-Xing Chen}
\affiliation{Institute for Quantum Science and Technology, National University of Defense Technology, Changsha Hunan 410073, China}
\affiliation{Hunan Key Laboratory of Mechanism and Technology of Quantum Information, Changsha Hunan 410073, China}

\begin{abstract}
    {Accurate thermometry of laser-cooled ions is crucial for the performance of the trapped-ions quantum computing platform. However, most existing methods face a computational exponential bottleneck. Recently, a thermometry method based on bichromatic driving was theoretically proposed by Ivan Vybornyi et al. to overcome this obstacle, which allows the computational complexity to remain constant with the increase of ion numbers. In this paper, we provide a detailed statistical analysis of this method and prove its robustness to several imperfect experimental conditions using Floquet theory. We then experimentally verify its good performance on a linear segmented surface-electrode ion trap platform for the first time. This method is proven to be effective from near the motional ground state to a few mean phonon numbers. Our theoretical analysis and experimental verification demonstrate that the scheme can accurately and efficiently measure the temperature in ion crystals.}
\end{abstract}

\date{June 4, 2024}
\maketitle

\address{}

\vspace{8mm}

\section{Introduction}\label{sec:introduction}
Surface ion traps are important development direction for scalable trapped-ion quantum computers~\cite{Mehta2016,Mehta2020,Ivory2021}. Due to the serious heating problem of surface traps, it is essential to develop thermometry methods characterized by straightforward experimental configuration, fast measurement speed, applicability in multi-ion scenarios, minimal computational resource consumption, and an extensive measurement range. Such methods will enable precise evaluations of cooling schemes and heating rates for these platforms.

From the perspective of parameter estimation, thermometric approaches near the motional ground state are categorized into two types: single-parameter and multi-parameter estimation schemes. Single-parameter estimation assumes a thermal distribution of phonons, requiring only the estimation of the average phonon number. Techniques in this category include fitting the time evolution of a red or blue sideband process~\cite{Poschinger2011}, blue-to-red sideband ratio~\cite{Leibfried2003}, coupling via a bichromatic laser field in a transverse mode~\cite{Ivanov2019}, and rapid adiabatic passage (RAP) techniques~\cite{Kirkova2021}. By contrast, multi-parameter estimation involves measuring the population of each Fock state within the phonon space, with methods such as red-sideband adiabatic evolution~\cite{An2014}, singular value decomposition (SVD) frequency-domain analysis~\cite{Meekhof1996}, composite pulses~\cite{Li2024}, and time-average measurement of multi-order sideband processes~\cite{Rasmusson2021}.

Commonly utilized techniques based on the evolution of red or blue sidebands in experiments face a serious challenge, i.e. exponential computational bottlenecks when dealing with a large number of ions. The red and blue sideband ratio technique becomes biased at high phonon numbers, rendering it unsuitable for high temperatures and necessitating modifications. Additionally, the RAP method, which maps phonon excitations to qubit excitations, is slow and thus unsuitable for surface trap platforms with high heating rates.

The bichromatic driving scheme, recently introduced in Ref. \cite{Vybornyi2023}, has the potential to overcome the computational exponential bottleneck. However, the analytical results are based on the first-order Lamb-Dicke approximation. In this paper, we extend the analytical expression for the excited state population to higher temperatures, based on their work. We find that the results in Ref. \cite{Vybornyi2023} remain accurate for estimating a higher mean phonon number. We further analyze the statistical characteristics of this method, and give an estimator that utilizes information from all qubits. Additionally, we explore solutions in the strong coupling regime using Floquet theory. We then experimentally validate the scheme on a linear segmented surface-electrode ion trap platform \cite{Ou2016,Xie2017}. We measure the cooling limits and heating rates of the center of mass (COM) mode using blue-sideband evolution, red-sideband evolution, and bichromatic driving methods. We also measure the cooling limit of the breathing mode. By comparing these results, we demonstrate the accuracy of the bichromatic driving scheme. Consequently, this method allows for the determination of the temperature of a motional mode in multi-ion systems without significant computational resources. Despite its advantages, the method has limitations, which we will discuss at the end of the paper.

The structure of this paper is as follows: Sec.~\ref{sec:theoryA} presents the model of bichromatic driving. Sec.~\ref{sec:theoryB} discusses the statistical analysis. Sec.~\ref{sec:experiment} gives the details of  our experimental procedure and provides a comprehensive description of the numerical processing involved. We conclude and provide future perspectives in Sec.~\ref{sec:conclude}.

\section{Theory of Bichromatic Driving Thermometry}\label{sec:theory}
\subsection{Model}\label{sec:theoryA}

We consider a quantum composite system described by two Hilbert subspaces: the qubit subspace and the harmonic oscillator (phonon) subspace. For trapped ions we consider here, these subspaces can be coupled using a sideband laser. Our ability is limited to detecting the qubit state population, rather than performing direct projective measurement on the phonon subspace. Consequently, the information from the harmonic oscillator must be transferred into the qubit subspace through a positive-operator valued measurement (POVM). We will first concentrate on the state of the harmonic oscillator.

Assuming that the phonon has a thermal population characterized by a single free parameter, the average phonon number $\bar{n}$,
\begin{align}
    P^{th}_{n}\left(\bar{n}\right)=\frac{\bar{n}^{n}}{\left(\bar{n}+1\right)^{n+1}},
\end{align}
and we can estimate its value directly by a certain technique which is a projection valued measurement (PVM). The quantum Fisher information (QFI) of $\bar{n}$ is~\cite{Kirkova2021}
\begin{align}
    F_Q\left(\bar{n}\right)=\frac{\bar{n}\sqrt{\frac{1}{\bar{n}}+1}}{2}\mathrm{ln}^2\left(\frac{1}{\bar{n}}+1\right).
    \label{eq:PVM}
\end{align}
QFI can be derived by optimizing the set of measurements for classical Fisher information (CFI), with the maximum CFI equating to QFI, i.e. $F_{C}\leq F_{Q}$. Consequently, a measurement scheme can be evaluated based on how closely its CFI approaches the QFI.

The model under consideration involves a red-detuned laser and a blue-detuned laser of the same equivalent strength and absolute detuning, which couple the qubits to a specific phonon mode. The Hamiltonian for our model is described by
\begin{align}
    H & =\sum_i h_i=-\sum_{i}\delta_i\ket{e}_{ii}\bra{e}\nonumber                                                                                             \\
      & +\frac{\Omega}{2}\left(\ket{e} _{ii}\bra{g}+\ket{g}_{ii}\bra{e}\right)\left[\hat{A_i}(\rme^{-\im\Delta_i t}+\rme^{\im\Delta_i t})+\right.\nonumber \\
      & \left.\hat{M}_i\rme^{-\im(\nu-\Delta_i)t}+\hat{M}_i^{\dagger}\rme^{\im(\nu-\Delta_i)t}\right],
\end{align}
where the frequency inhomogeneity of ions $\delta_i$ is caused by inhomogeneous magnetic field and $i$ is the index of the ion, $\nu$ is the phonon angular frequency, $\Delta$ is the laser detuning to the carrier resonant frequency, $\Omega$ is the coupling strength between a qubit and the phonon, $\ket{g}$ and $\ket{e}$ denote the qubit's ground and excited states. We use $\hat{a}$ to denote the phonon annihilation operator and $\ket{n}$ to denote the phonon number state. The  operators $\hat{A}_i$, and $\hat{M}_i$ are defined as $\hat{A}_i\equiv\sum_n \rme^{-\frac{\eta_i^2}{2}} \mathcal{L}^0_{n}(\eta_i^2)\ket{n}\bra{n}$ and $\hat{M}\equiv \sqrt{n+1} \eta_i \rme^{-\frac{\eta_i^2}{2}} \mathcal{L}_{n}^1\left(\eta_i^2\right)\ket{n}\bra{n+1},\ n=0,1,2,3,\cdots$, where $\eta_i$ is the Lamb-Dicke parameter for a selected vibrational mode on the $i$th ion, and $\mathcal{L}^{\alpha}_{n}$ is the generalized Laguerre polynomial~\cite{Wineland1979,Rasmusson2021}. In weak coupling regime,$\hat{A}_i\approx I$ and $\hat{M}_i\approx \hat{a}$.

The Hamiltonian described has several characteristics: Firstly and most importantly, each $h_i$ commutes with the others, thereby simplifying the computational process significantly. This commutation allows for the analysis of individual qubits and phonon modes, rather than requiring a comprehensive examination of all qubits simultaneously. Secondly, under conditions of weak coupling strength, equal laser detuning to the phonon frequency, and uniform Zeeman splitting, the Hamiltonian is represented as $h_i=\frac{\eta_i\Omega}{2}\left(\ket{g}_{ii}\bra{e}+\ket{e}_{ii}\bra{g}\right)(\hat{a}+\hat{a}^{\dagger})$, as demonstrated in Ref.~\cite{Vybornyi2023}. Thirdly, experimental conditions with strong laser intensities may lead to significant AC Stark effects, and non-resonant carrier excitations cannot be disregarded. Additionally, if the Zeeman splitting is not uniform, Floquet theory becomes necessary to analytically describe the system's evolution, which will be ~\cite{Mikami2016,Goldman2014}
\begin{align}
    U=\prod_i{U_i},
\end{align}
where
\begin{align}
    U_i=\mathrm{e}^{-\mathrm{i}K\left(t_f\right)}\mathrm{e}^{-\mathrm{i}\left(t_f-t_i\right)H_{eff}}\mathrm{e}^{\mathrm{i}K\left(t_i\right)},
    \label{eq:ueff}
\end{align}
\begin{align}
    H_{eff}=-\delta_i\ket{e}_{ii}\bra{e}+\frac{\Omega}{2}\left(\ket{e}_{ii}\bra{g}+\ket{g}_{ii}\bra{e}\right)(\hat{M}_i+\hat{M}_i^{\dagger}),
    \label{eq:heff}
\end{align}
and
\begin{align}
    K(t)=\frac{\Omega}{\im 2\nu}\left(\ket{e}_{ii}\bra{g}+\ket{g}_{ii}\bra{e}\right)(\rme^{-\im\nu t}+\rme^{\im\nu t}).
    \label{eq:k}
\end{align}
From Eq.~\eqref{eq:k} we can find that, in this situation, if we choose discrete measuring time points at $k\frac{2\pi}{\nu},\ k=1,2,3,\cdots$, the evolution is as the same of the weak coupling regime. The derivation of above evolution can be found in Appendix.~\ref{sec:vv}.

We generalize the analytical expression of the excited state population to a higher temperature, based on the previous works in weak coupling regime, as
\begin{align}\label{eq:pe}
     & P^i_e\left(\bar{n},t\right)=\frac{1}{2}\left[1-\mathrm{e}^{-\eta_i^2\Omega^2t^2\left(\frac{1}{2}+\bar{n}\right)}\right]\nonumber                                                \\
     & -\frac{1}{2}\eta_i^4\Omega^2t^2\left(\bar{n}+1\right)\left[2-\eta_i^2\Omega^2t^2\left(1+\bar{n}\right)\right]\mathrm{e}^{-\eta_i^2\Omega^2t^2\left(\frac{1}{2}+\bar{n}\right)}.
\end{align}
The derivation of Eq.~\eqref{eq:pe} is presented in Appendix~\ref{sec:pe}. In a wide range of $\bar{n}$, Eq.~\eqref{eq:pe} can be reduced to
\begin{align}\label{eq:pie}
    P^i_e\left(\bar{n},t\right)=\frac{1}{2}\left[1-\mathrm{e}^{-2\left(\eta_i\frac{\Omega}{2}t\right)^2\left(1+2\bar{n}\right)}\right].
\end{align}
Consequently, the analytical results of Eq.~\eqref{eq:pie} applicable in the near motional ground state remain valid at higher temperatures. Both analytical and numerical verifications confirm that the expansion maintains sufficient accuracy for $\bar{n}<20$.

\subsection{Temperature estimation}\label{sec:theoryB}
In this subsection, we give the statistics analysis of bichromatic driving method for estimating $\bar{n}$ of the ion crystal.

When conducting $N$ measurements, the variance $\Delta\bar{n}^2$ is constrained by the Cramér-Rao bound (CRB)
\begin{align}
    \Delta\bar{n}^2\geq\frac{1}{NF(\bar{n})}.
\end{align}
Therefore, a higher Fisher information value in thermometry corresponds to increased accuracy.

\begin{figure}[htbp]
    \centering
    \includegraphics[width=\columnwidth]{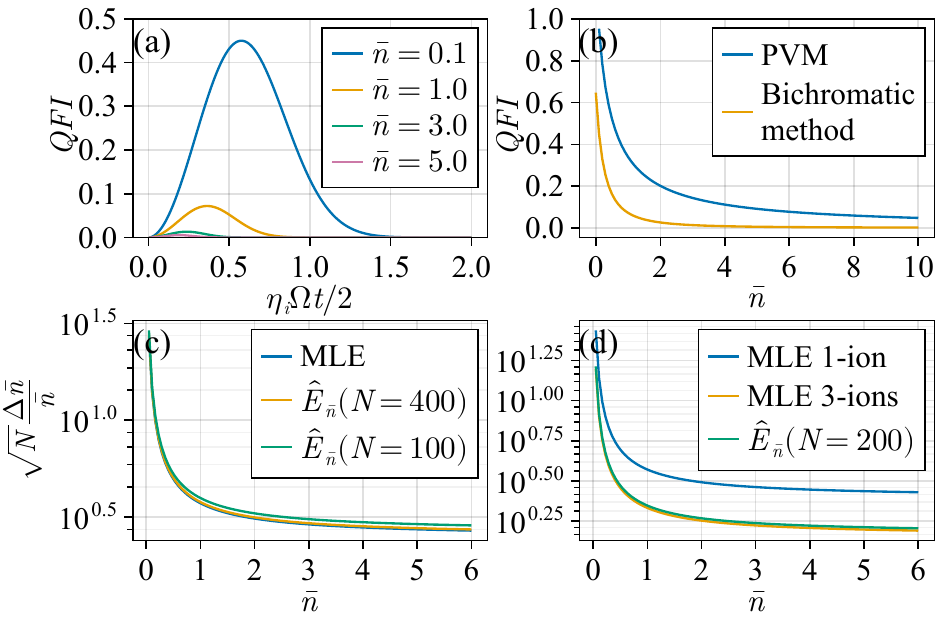}
    \caption{(a)The QFI of bichromatic driving thermometry depends on both the mean photon number, $\bar{n}$, and the time, $t$. (b) A comparison of the Fisher information as expressed in Eq.~\eqref{eq:PVM} and~\eqref{eq:fisher_s}. (c) A comparison of the standard deviation derived from the MLE using the CRB and the estimator described in Eq.~\eqref{eq:est}. With 400 detections, the variance of the estimator approaches that of the CRB.  (d) A comparison of standard deviation using MLE with single qubit (MLE 1-ion), 3 qubits (MLE 3-ions) and Eq.~\eqref{eq:e3}, which indicates that simultaneously using all the state information can significantly improve measuring accuracy.}
    \label{fig:sa}
\end{figure}

The QFI of bichromatic thermometry is equivalent to its CFI due to the vanishing off-diagonal elements of the qubit's reduced density matrix, $P_{eg}=P_{ge}=0$. A detailed derivation can be found in Appendix~\ref{sec:CQ}. The Fisher Information for bichromatic driving thermometry is
\begin{align}\label{eq:fisher_s}
    F_i\left(\bar{n},t\right)=\frac{\left(\partial_{\bar{n}}P^i_e\right)^2}{P^i_e}+\frac{\left(\partial_{\bar{n}}P^i_g\right)^2}{P^i_g}=\frac{16\left(\eta_i\frac{\Omega}{2}t\right)^4}{\mathrm{e}^{4\left(\eta_i\frac{\Omega}{2}t\right)^2\left(2\bar{n}+1\right)}-1},
\end{align}
which is a function of $\bar{n}$ and $t$. In Fig.~\ref{fig:sa}(a), we show that different $\bar{n}$ have different optimal measuring time $t$. Fig.~\ref{fig:sa}(b) shows a comparison of Fisher information of Eq.~\eqref{eq:PVM} and Eq.~\eqref{eq:fisher_s}.
The statistical analysis indicates that the variance of an estimate depends upon the chosen estimator; the optimal estimator that achieves the CRB is the Maximum Likelihood Estimator (MLE). However, in practical estimation processes, MLE is rarely selected due to its complexity. Instead, a simpler expression is often preferred. In this instance, we adopt Eq.~\eqref{eq:est} as our estimator,
\begin{align}
    \hat{\bar{n}}=-\frac{\mathrm{ln}\left(1-2P^i_e\right)}{\left(\eta_i\Omega t\right)^2}-\frac{1}{2}.
    \label{eq:est}
\end{align}

We employ the electronic shelving method to detect the state of qubits \cite{Nagourney1986}. The number of qubits in state $\ket{e}$ with $N$ detections follows a binomial distribution. As the number of detections becomes large, the distribution of the population $p$ of $\ket{e}$ approximates a normal distribution \cite{Shiryaev1996},
\begin{align}
    f_{N\rightarrow\infty}\left(\hat{p}\right)=\frac{1}{\sqrt{2\mathrm{\pi}}\sigma_s}\mathrm{e}^{-\frac{\left(\hat{p}-p\right)^2}{2\sigma_s^2}}, 
\end{align}
where $\sigma_s=\frac{p\left(1-p\right)}{N}$, $\hat{p}$ is the estimate of $\ket{e}$'s population, and $p$ is the true value.
Using this estimator, the estimate bias is
\begin{align}
    \delta\bar{n}=\langle\hat{\bar{n}}\rangle-\bar{n}\approx-\frac{P^i_e\left(1-P^i_e\right)\left(2\bar{n}+1\right)}{N\left(1-2P^i_e\right)^2\mathrm{ln}\left(1-2P^i_e\right)},
\end{align}
and the variance is
\begin{align}
    {\Delta\bar{n}}^2=\langle\left(\hat{\bar{n}}-\bar{n}\right)^2\rangle\approx\frac{P^i_e\left(1-P^i_e\right)\left(2\bar{n}+1\right)^2}{N\left(1-2P^i_e\right)^2\mathrm{ln}^2\left(1-2P^i_e\right)}.
\end{align}

In our experiment, we possess the capability to simultaneously read the states of all qubits. Consequently, by utilizing the states of $M$ qubits rather than a single one, we can obtain a more accurate estimate. Therefore, the QFI becomes
\begin{align}
    F\left(\bar{n},t\right)=\sum^{2^M}_{k=1}\frac{\left(\partial_{\bar{n}}P_k\right)^2}{P_k};P_k=\prod^M_{i=1}P^i_l\left(\bar{n},t\right),l=\ket{e},\ket{g}.
\end{align}
We employ the estimator as the following type: 
\begin{align}
    \hat{E}_{\bar{n}}=\mathop{\arg\min}\limits_{\bar{n}}\sum_{i=1}^{M}\frac{(\hat{\bar{n}}_i-\bar{n})^2}{{\Delta\bar{n}_i}^2}.
    \label{eq:e3}
\end{align}
In Figs.~\ref{fig:sa} (c) and (d), we present a comparison of the standard deviations using MLE and Eq.~\eqref{eq:est} /Eq.~\eqref{eq:e3}. The results indicate that using the state information from all qubits simultaneously can significantly enhance measurement accuracy.

\section{Experimental Results}\label{sec:experiment}
To verify the accuracy of bichromatic driving thermometry, we measured the cooling limit and heating rate in our linear segmented surface-electrode ion trap platform using three methods: blue-sideband evolution fitting, red-sideband evolution fitting, and bichromatic driving thermometry. The results indicate that the red- and blue-sideband evolution fitting methods can achieve high accuracy due to the large volume of measurement data they employ. We point that these methods have sufficient accuracy to serve as benchmarks for evaluating the new bichromatic driving thermometry.

\begin{figure}[htbp]
    \centering
    \includegraphics[width=\columnwidth]{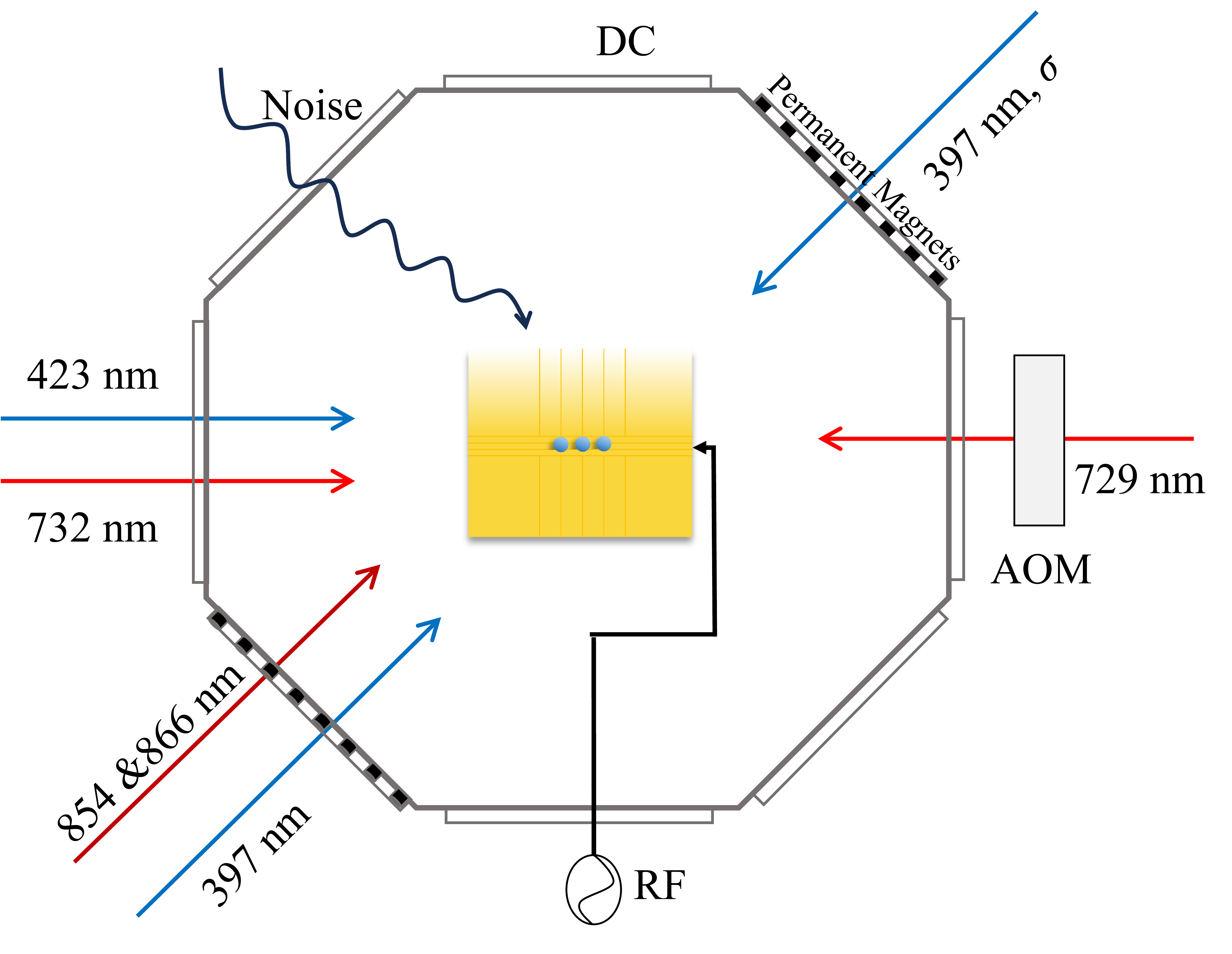}
    \caption{Schematic diagram of the experimental platform.}
    \label{fig:platform}
\end{figure}

We placed a surface-electrode ion trap in a vacuum chamber to confine $\rm{^{40}Ca^+}$ ions. Fig.~\ref{fig:platform} illustrates the laser setup in our platform. Lasers at 423 nm and 732 nm are used for ionizing the atoms, while lasers at 854 nm, 866 nm, and 397 nm are employed for state initialization. The 397 nm laser is also used for electronic shelving detection of the qubit state, and the 729 nm laser couples the axial phonon mode and qubits, encoding the temperature information of a phonon mode into the qubit. Due to the uncertainty of the heating rate of the surface-electrode ion trap, we actively introduced noise into the system to simulate heating \cite{Johnson2015}. An EMCCD camera is used to collect fluorescence from individual ions simultaneously during the electronic shelving detection process.

\begin{figure}[htbp]
    \centering
    \includegraphics[width=\columnwidth]{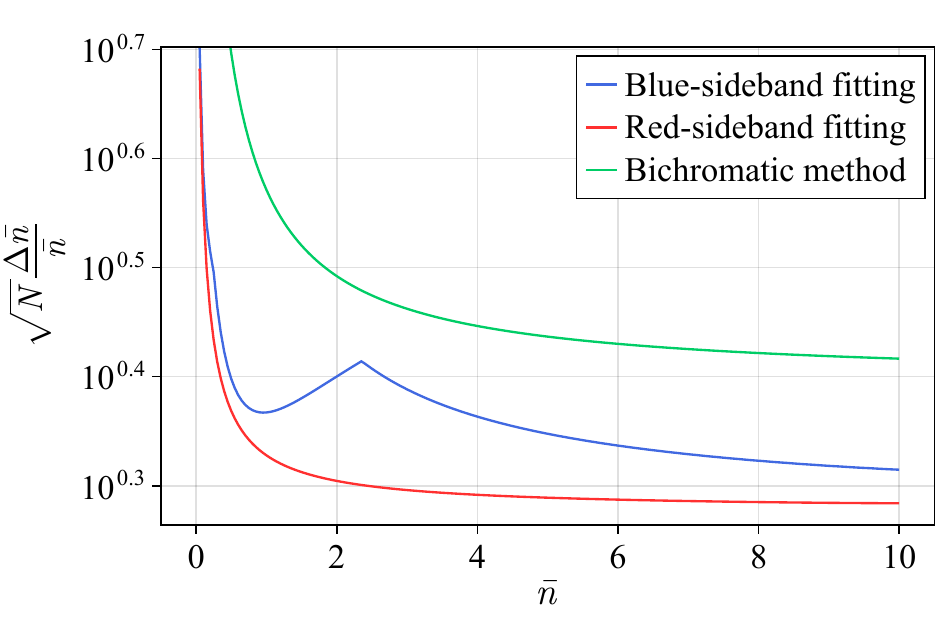}
    \caption{Minimum standard deviations bounded by the Cramér-Rao Bound (CRB) of blue-sideband, red-sideband, and bichromatic driving thermometry vary with $\bar{n}$ when considering the temperature measurement of a single ion at the optimal time point.}
    \label{fig:brb}
\end{figure}

Furthermore, we provide a comparison of three methods in Fig.~\ref{fig:brb}. When using bichromatic, red-sideband, or blue-sideband excitation for thermometry, as previously mentioned, their quantum Fisher information is a function of $\bar{n}$ and evolution time $t$. Fig.~\ref{fig:brb} shows the standard deviation changes with temperature at their respective optimal time points. This result indicates that estimating temperature by measuring at an optimal time point theoretically yields the highest accuracy. In the experiment, we adjusted the specific measurement process according to the characteristics of different schemes. For red or blue sideband thermometry, we typically fit the evolution of the qubit  (which has $m$ data points) to estimate $\bar{n}$ rather than using a single time point, in which case $\hat{\bar{n}}=\arg\min\limits_{\bar{n}}\sum_{i=1}^{m}\frac{(\bar{n}-\bar{n}_i)^2}{\sigma_i^2}$. However, for bichromatic driving thermometry, we continue to use a single time point.

We initially measure the heating rate of the center-of-mass (COM) phonon mode using both red-sideband and blue-sideband thermometry. Specifically, we monitor the evolution of the qubit excitation during a blue-sideband process. The evolution time ranges from $0$ to $200 \mu s$, in intervals of $2 \mu s$. The subsequent step involves data processing.

This is an optimization problem aiming at determining the value of $\bar{n}$ that best fits the evolution of the sideband process. The loss function we employ is:
\begin{align}
    loss\left(\bar{n}\right)=\sum^M_{i=1}\sum^m_{j=1}\frac{\left[P^i\left(\bar{n},t_j\right)-x_{i,j}\right]^2}{\sigma_{i,j}^2},
    \label{eq:loss}
\end{align}
where $M$ is the number of ions, $j$ is the index of discrete detection time, $P$ is the population of qubit excitation calculated numerically, $x$ comes from the experimental data and $\sigma^2$ is the variance of $x$, and $\hat{\bar{n}}=\mathop{\arg\min}\limits_{\bar{n}}\ loss(\bar{n}) $.
In Eq.~\eqref{eq:loss}, the state of all qubits is considered when fitting $\bar{n}$, allowing us to fully utilize the state data obtained from EMCCD readings. The variance of this estimator is derived from the theory of error estimation in non-linear least squares data analysis \cite{Cline1970},
\begin{align}
    {\Delta\bar{n}}^2\approx\left[\sum^{M,m}_{i,j=1}\left(\frac{A_{i,j}}{\sigma_{i,j}}\right)^2\right]^{-1}\frac{loss\left(\bar{n}\right)}{M m-1}F\left(1,M m-1;1-\beta\right),
\end{align}
where
\begin{align}
     & A_{i,j}=\frac{\partial P^i\left(\bar{n},t_j\right)}{\partial\bar{n}}=\frac{\partial\vec{P}_{thermal}(\bar{n})}{\partial\bar{n}}.\vec{P}^i(t_j),\nonumber \\
     & \vec{P}_{thermal}(\bar{n})=\left[P^{th}_{0}(\bar{n}),P^{th}_{1}(\bar{n}),P^{th}_{2}(\bar{n}),\cdots\right], \nonumber                                    \\
     & \vec{P}^i(t_j)=\left[p^i_0(t_j),p^i_1(t_j),p^i_2(t_j),\cdots\right]\nonumber,
\end{align}
$\sigma_{i,j}^2$ is the variance of the experiment data of the $i$th ion at $j$th time point, $F(1,M m-1)$ is the statistical $F$ distribution and $\beta=0.317$; $p^i_n(t_j)$ in the last vector is the population of $|e\rangle_i|n\rangle$ at time $t_j$.
We compute the partial derivative using $Zygote$~\cite{Innes2018} automatic differentiation package in $Julia$~\cite{Bezanson2012}. And we use $Optimal$~\cite{KMogensen2018} package to search the minimum and minimizers of the loss function. The results are plotted using $CairoMakie$~\cite{Danisch2021} package. 

To improve the accuracy of the estimation, it is essential to emphasize one particular issue. Prior to performing the sideband process, we must experimentally measure the Rabi frequency of the coupling ($\Omega$) and the phonon mode frequency ($\nu$) of interest. Due to the AC Stark shift induced by non-resonant carrier excitation, the first-order sideband frequency does not equal the mode frequency, necessitating knowledge of the sideband laser detuning ($\Delta$). The evolution is calculated using the Hamiltonian $\nu \hat{a} \hat{a}^{\dagger}-\Delta \sum_{i}\ket{e}_{ii}\bra{e}+\sum_{i}\frac{\Omega}{2}\left(\ket{e}_{ii}\bra{g}\rme^{\im \eta_i (\hat{a}+\hat{a}^{\dagger})}+h.c.\right)$. With these parameters, we can generate a file containing the evolution of the excited state population, sweeping the initial state from $\ket{g}\ket{0}$ to $\ket{g}\ket{n}$. This numerical data allows us to fit the evolution and determine $\bar{n}$.

\begin{figure}[htbp]
    \centering
    \includegraphics[width=\columnwidth]{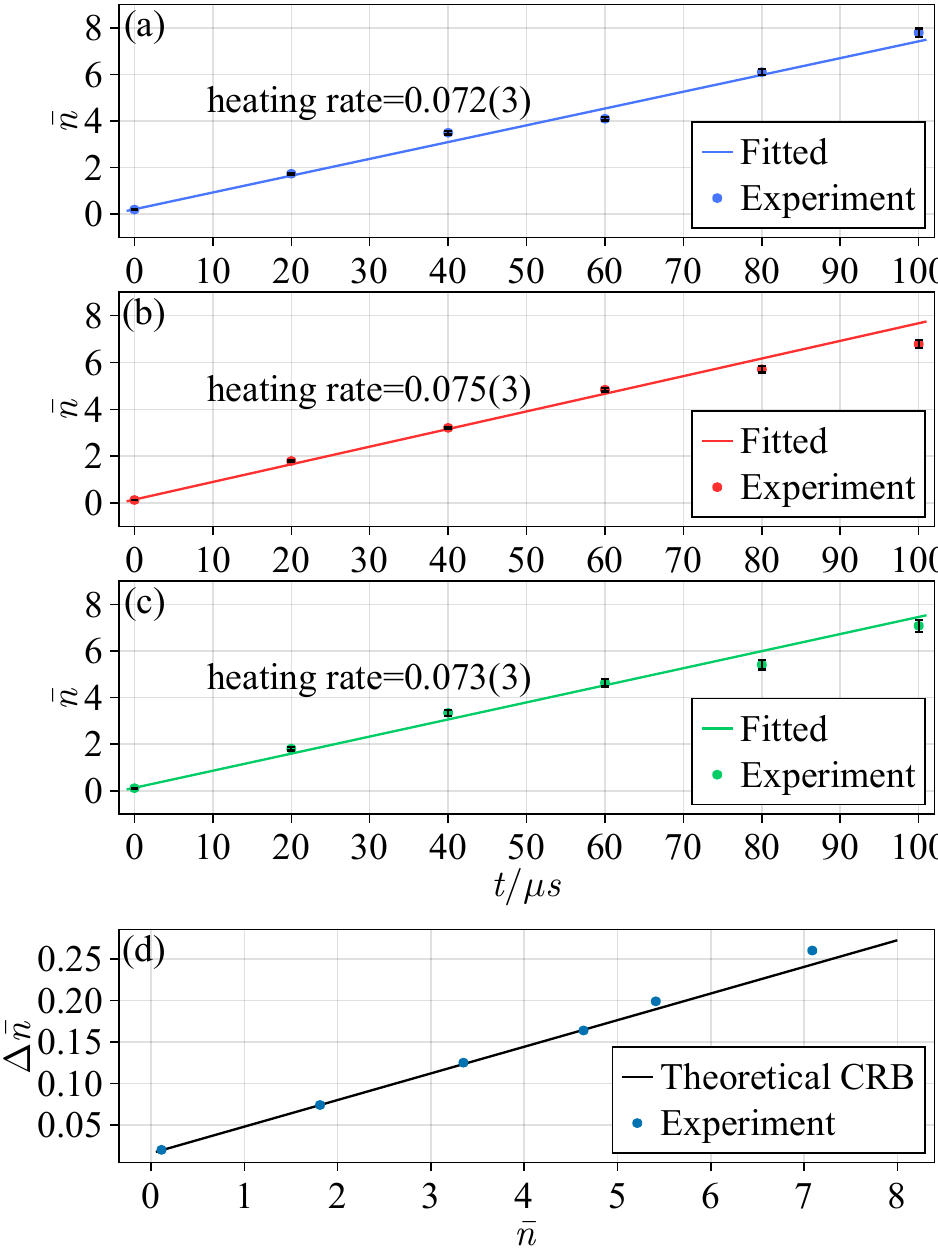}
    \caption{Subfigures (a), (b), and (c) respectively present the experimental data for the heating rate, including error bars, and the corresponding fitting lines obtained through blue-sideband, red-sideband, and bichromatic driving thermometries. In (d), the black line represents the theoretical bound of the change in $\Delta\bar{n}$ relative to $\bar{n}$, while the blue dots indicate the experimental data for $\Delta\bar{n}$.}
    \label{fig:comp3}
\end{figure}

When using bichromatic thermometry, we conduct 2000 repeated detections at an optimal time point, rather than fitting the evolution as done in sideband processes. In our experiment, the initial temperature is unknown prior to measurement, thus preventing us from specifying an accurate detection time point. By substituting Eq.~\eqref{eq:pie} into Eq.~\eqref{eq:fisher_s}, we derive the following equation:
\begin{align}
    F_i\left(P^i_e,\bar{n}\right)=\frac{4}{(2\bar{n}+1)^2}\frac{\mathrm{ln}^2\left(1-2P^i_e\right)}{\left(1-2P^i_e\right)^{-2}-1}.
\end{align}
From this equation, we observe that the maximum point of $F$ at different values of $\bar{n}$ corresponds to the same $P_e$. It is straightforward to calculate that $P_{e}\approx 0.274618$. For experimental convenience, we initially perform an evolution time scan, repeating the measurement 100 times at each point. We then identify the time point where the population of the excited state is approximately 0.27. This time is chosen as the optimal time point, at which we repeat the measurement 2000 times to estimate $\bar{n}$.

In Fig.~\ref{fig:comp3}, we present the results of our experiments. The heating rates obtained using three different thermometry methods are as follows: blue-sideband thermometry yielded $0.072(3)/\mu s$, red-sideband thermometry yielded $0.075(3)/\mu s$, and bichromatic driving thermometry yielded $0.073(3)/\mu s$. The close agreement among these heating rates corroborates the accuracy of each method. The error bars in our experiments align with the predictions shown in Fig.~\ref{fig:brb}, indicating that the variances for red-sideband and blue-sideband thermometry are similar, and the variance for bichromatic driving thermometry is also minimal. Another significant finding, depicted in Fig.~\ref{fig:comp3}(d), is that the standard deviation of the experimental data for bichromatic driving thermometry nearly reaches the CRB of the theoretical prediction. Thus, we have validated the accuracy of bichromatic driving thermometry both theoretically and experimentally.

We also tested bichromatic driving thermometry in the axial breathing mode. The cooling limit measured by blue-sideband thermometry is 0.59 with a standard deviation of 0.03. In contrast, using bichromatic driving thermometry, we obtained a cooling limit of 0.47 with a variance of 0.03. Therefore, this thermometry method remains effective for measuring non-COM modes.

\section{DISCUSSION AND CONCLUSIONS}\label{sec:conclude}
In this paper, we extend the thermometry measurement regime discussed in Ref.~\cite{Vybornyi2023} to $\bar{n} < 20$, demonstrating that the evolution expression for the excited state population remains valid. We then present a detailed statistical analysis of this bichromatic driving thermometry. The accuracy of our method is verified by comparing it with data from three different thermometric approaches. In our experiments, we use an EMCCD to detect the state of individual qubits, allowing us to fit or calculate $\bar{n}$ using data from all qubits. The experimental results show that the accuracy of bichromatic driving thermometry approaches the theoretical CRB. We also evaluate its applicability in non-COM motional modes, concluding that this method is computationally efficient, user-friendly, and accurate.

Although all three methods are applicable and accurate on a small scale, the red- or blue-sideband method will encounter an exponential bottleneck problem when the scale becomes larger, while the bichromatic driving method will not. This will be the focus of our future work.

However, we also identified some limitations of bichromatic driving thermometry during our experiments. In Eq.~\eqref{eq:pie}, the parameters $\Omega$, $t$, and $\bar{n}$ are multiplicatively related, necessitating precise measurement of $\Omega$ and accurate control of $t$. Some anomalies, such as delays in RF circuits, may go undetected if only single time-point measurements are considered. 

\begin{acknowledgments}
    This work is supported by National Natural Science Foundation of China under Grants No.12074433, 12174447, 12174448, 12204543.
\end{acknowledgments}

\appendix
\section{VAN VLECK EXPANSIONS}\label{sec:vv}
Given a time-periodic driving system with multiple frequencies, all sharing a common divisor $\omega$, a Fourier transform can be applied. The Fourier coefficient for each component is:
\begin{align}
    H_{n}=\frac{\omega}{2\pi}\int^{\frac{2\pi}{\omega}}_0 dt\mathrm{e}^{\mathrm{i}n\omega t}H(t),
\end{align}
where $H(t)$ is the time periodic Hamiltonian.
The evolution of this system can be decomposed into a van Vleck degenerate perturbation form, 
\begin{align}
    U_i=\mathrm{e}^{-\mathrm{i}K\left(t_f\right)}\mathrm{e}^{-\mathrm{i}\left(t_f-t_o\right)H_{eff}}\mathrm{e}^{\mathrm{i}K\left(t_o\right)},
\end{align}
where
\begin{align}
     & H_{eff}=\sum^{\infty}_{n=0}H^{(n)},\nonumber                                          \\
     & H^{(0)}=H_0,\nonumber                                                                 \\
     & H^{(1)}=\sum_{m\neq0}\frac{[H_{-m},H_{m}]}{2m\omega},\nonumber                        \\
     & H^{(2)}=\sum_{m\neq0}\frac{[[H_{-m},H_0],H_m]}{2m^2\omega^2},\nonumber                \\
     & \qquad \quad +\sum_{m\neq0}\sum_{n\neq0,m}\frac{[[H_{-m},H_{m-n}],H_n]}{3mn\omega^2},
\end{align}
and
\begin{align}
     & K(t)=\sum^{\infty}_{n=0}K^{(n)}(t),\nonumber                                                             \\
     & K^{(1)}=\mathrm{i}\sum_{m\neq0}\frac{H_m}{m\omega}\mathrm{e}^{\mathrm{i}m\omega t},\nonumber             \\
     & K^{(2)}=-\im\sum_{m\neq0}\sum_{n\neq0,m}\frac{[H_n,H_{m-n}]}{2mn\omega^2}\rme^{-\im m\omega t},\nonumber \\
     & \qquad \quad-\im \sum_{m\neq0}\frac{[H_m,H_0]}{m^2\omega^2}\rme^{-\im m\omega t}.
\end{align}
When the coupling is weak, the Zeeman splitting is homogeneous, and the laser detuning matches the selected phonon mode frequency, the Hamiltonian of a single qubit interacting with the phonon mode can be described as follows:
\begin{align}
    H_i &= \delta_i\ket{e}_{ii}{\bra{e}}
      +\frac{\Omega}{2}\left(\ket{e}_{ii}\bra{g}+\ket{g}_{ii}\bra{e}\right)\left[I(\rme^{-\im\Delta_i t}+\rme^{\im\Delta_i t})\right.\nonumber \\
      & +\left.\hat{M_i}+\hat{M}_i^{\dagger}\right].
      \label{eq:hw}
\end{align}
We can easily get Eq.~\eqref{eq:ueff}, Eq.~\eqref{eq:heff} and Eq.~\eqref{eq:k} by substituting Eq.~\eqref{eq:hw} into van Vleck expansion.
\section{DERIVATION OF EXCITED STATE POPULATION}\label{sec:pe}

We give the derivation process in this appendix. We show an easy case first. Considering that a Hamiltonian of a single ion is $h_i=\frac{\eta_i\Omega}{2}\left(\ket{g}_{ii}\bra{e}+\ket{e}_{ii}\bra{g}\right)(\hat{a}+\hat{a}^{\dagger})$, the evolution of the system is 
\begin{align}
    \mathrm{e} & ^{-\im\left[\frac{\eta_i \Omega}{2}(|e\rangle_{ii}\langle g|+| g\rangle_{ii}\langle e|)(\hat{a}+\hat{a}^{\dagger})\right] t}|g\rangle_{ii}\langle g| \nonumber                                                 \\
               & \otimes\left[\sum_n p_n|n\rangle\langle n|\right] \mathrm{e}^{\im\left[\frac{\eta_i \Omega}{2}\left(|e\rangle_{ii}\langle g|+| g\rangle_{ii}\langle e|\right)(\hat{a}+\hat{a}^{\dagger})\right] t}\nonumber \\
    =          & \sum_n p_n \mathrm{e}^{-\im\frac{\eta_i \Omega}{2}\left(|e\rangle_{ii}\langle g|+\mid g\rangle_{ii}\langle e|\right)(\hat{a}+\hat{a}^{\dagger}) t}|g\rangle_i|n\rangle\nonumber                \\
               & \langle n|_i\langle g| \mathrm{e}^{\im\frac{\eta_i \Omega}{2}\left(|e\rangle_{ii}\langle g|+\mid g\rangle_{ii}\langle e|\right)(\hat{a}+\hat{a}^{\dagger}) t}.
\end{align}
The first key step of the derivation is to represent the qubit state in the eigenbasis of $(\ket{g}_{ii}\bra{e}+\ket{e}_{ii}\bra{g})$, such that
\begin{align}
     & \mathrm{e}^{-\im\frac{\eta_i \Omega}{2}\left(|e\rangle_{ii}\langle g|+| g\rangle_{ii}\langle e |\right)(\hat{a}+\hat{a}^{\dagger}) t}|g\rangle_i|n\rangle\nonumber                     \\
     & =\frac{1}{\sqrt{2}}|+\rangle_i \rme^{\frac{\im \eta_i \Omega t}{2}(\hat{a}+\hat{a}^{\dagger})}|n\rangle-\frac{1}{\sqrt{2}}|-\rangle_i \rme^{\frac{-\im \eta_i \Omega t}{2}(\hat{a}+\hat{a}^{\dagger})}|n\rangle.
\end{align}
Then, we have
\begin{align}
    P_e^i(\bar{n}, t)=\frac{1}{2}\left[1-\sum_n \frac{\bar{n}^n}{(\bar{n}+1)^{n+1}}\langle n|\right.\nonumber \\
    \left.\frac{\rme^{\im \eta_i \Omega t(\hat{a}+\hat{a}^{\dagger})}+\rme^{-\im \eta_i \Omega t(\hat{a}+\hat{a}^{\dagger})}}{2}| n\rangle\right].
\end{align}
Using  Baker-Campbell-Hausdorff (BCH) formula, we get
\begin{align}
    \langle n|e^{\im \eta_i \Omega t\left(\hat{a}+\hat{a}^{\dagger}\right)}| n\rangle & =\langle n|\rme^{-\alpha \prime a} \rme^{\alpha a^{\dagger}} \rme^{\frac{|\alpha|^2}{2}}| n\rangle\nonumber \\
    & =\rme^{\frac{|\alpha|^2}{2}}\left[1+\frac{\alpha^2}{(1!)^2}(n+1)+\right.\nonumber                           \\
    & \left.\frac{\alpha^4}{(2!)^2}(n+1)(n+2)+\cdots\right],
    \label{eq:b5}
\end{align}
where $\alpha=i \eta_i \Omega t$. After this, we get
\begin{align}
    P_e^i(\bar{n}, t)= & \sum_{n=0} p_n\left(1+\frac{\alpha^2}{(1!)^2}(n+1)+\right.\nonumber \\
                     & \left.\frac{\alpha^4}{(2!)^4}(n+1)(n+2)+\cdots\right) ,\nonumber    \\
\end{align}
where $p_n=\left(\frac{n}{\bar{n}+1}\right) p_0, p_0=\frac{1}{\bar{n}+1}$. The summation operations in Eq.~\eqref{eq:b5} can be done with the following transformation:
\begin{align}
     & \sum_{n=0} p_n(n+1)=\frac{1}{\bar{n}+1} \sum_{n=0}(n+1)\left(\frac{\bar{n}}{\bar{n}+1}\right)^n\nonumber                                       \\
     & =\frac{1}{\bar{n}+1} \frac{\partial}{\partial k} \sum_{n=0} k^{n+1}=\frac{1}{\bar{n}+1} \frac{\partial}{\partial k}\left(\frac{1}{1-k}\right),\nonumber
\end{align}
\begin{align}
     & \sum_{n=0} p_n(n+1)(n+2)=\frac{1}{\bar{n}+1} \frac{\partial^2}{\partial k^2}\left(\frac{1}{1-k}\right),\nonumber \\
     & \sum_{n=0} p_n(n+1)(n+2)(n+3)=\frac{1}{\bar{n}+1} \frac{\partial^3}{\partial k^3}\left(\frac{1}{1-k}\right),
     \label{eq:b6}
\end{align}
where $k=\frac{\bar{n}}{\bar{n}+1} $.
According to Eq.~\eqref{eq:b6}, we have
\begin{align}
    \langle n|e^{-\alpha^{\prime} a} e^{\alpha a^{\dagger}}| n\rangle= & 1+\frac{\alpha^2}{(1!)^2} \frac{1}{\bar{n}+1} \frac{\partial}{\partial k}\left(\frac{1}{1-k}\right)+\nonumber \\
    & \frac{\alpha^4}{(2!)^2} \frac{1}{\bar{n}+1} \frac{\partial^2}{\partial k^2}\left(\frac{1}{1-k}\right)+\cdots,\nonumber
\end{align}
\begin{align}
    \frac{1}{\bar{n}+1} \frac{\partial^m}{\partial k^m}\left(\frac{1}{1-k}\right)=m!(1+\bar{n})^m.
\end{align}
Finally, after a few easy derivations, we obtain the expression of $P^i_e(\bar{n},t)$ whose initial phonon thermal state is in near ground motional state,
\begin{align}
    P_e^i(\bar{n}, t)=\frac{1}{2}\left[1-\rme^{-2\left(\eta_i \frac{\Omega}{2} t\right)^2(2 \bar{n}+1)}\right].
\end{align}
When measuring temperatures slightly above the ground state, it is essential to consider higher-order terms in the exponential expansion rather than just the first order. Assuming weak coupling and that the laser resonates with the first-order sideband, the Hamiltonian is given by $h_i = \frac{\eta_i \Omega}{2} \left( \ket{g}_{ii}\bra{e} + \ket{e}_{ii}\bra{g} \right) (\hat{M}_i +\hat{M}_i^{\dagger}) $. Then, 
\begin{align}
    P_e^i(\bar{n}, t)=\frac{1}{2}\left[1-\sum_n \frac{\bar{n}^n}{(\bar{n}+1)^{n+1}}\langle n|\right.\nonumber \\
    \left.\frac{\rme^{\im \Omega t\left(\hat{M}_i+\hat{M}_i^{\dagger}\right)}+\rme^{-\im \Omega t\left(\hat{M}_i+\hat{M}_i^{\dagger}\right)}}{2}| n\rangle\right].
\end{align}
It is easy to verify that $\hat{M}_i^{\dagger},[\hat{M}_i,\hat{M}_i^{\dagger}]]=[\hat{M}_i,[\hat{M}_i,\hat{M}_i^{\dagger}]]=0$. So we can still use BCH formula as before. The specific process is omitted here. The key step to obtain the analytical expression is to find a suitable approximation of elements of operator $\hat{M}_i$$(\hat{M}_i^{\dagger})$. After trying assisted by numerical calculation, we choose
\begin{align}
    \hat{M}_i[n, n+1]=\hat{M}_i^{\dagger}[n+1, n] \approx f(n)\nonumber \\
    =\eta_i \sqrt{n+1}-\frac{\eta_i^3}{2} \frac{n(n+1)}{\sqrt{n+1}}.
\end{align}
The rest of the derivations are similar to the previous ones. We finally get expression Eq.~\eqref{eq:pe}. Thus we can conclude by our analytical derivation that the accuracy of Eq.~\eqref{eq:pie} is still enough for a higher temperature.

\section{CFI=QFI}\label{sec:CQ}

According to the characteristic of trace and $\rme^{\hat{A}}\hat{B}\rme^{-\hat{A}}=\hat{B}+[\hat{A},\hat{B}]+\frac{1}{2!}[\hat{A},[\hat{A},\hat{B}]]$, we can easily get
\begin{align}
     & P_{eg}^i+P_{ge}^i=Tr\left[(\ket{g}_{ii}\bra{g}\otimes\rho_{th})\rme^{\im \frac{\eta_i\Omega}{2}\left(\ket{g}_{ii}\bra{e}+\ket{e}_{ii}\bra{g}\right)(\hat{a}+\hat{a}^{\dagger})t}\right.\nonumber \\
     & \left.(\ket{e}_{ii}\bra{g}+\ket{g}_{ii}\bra{e})\rme^{-\im \frac{\eta_i\Omega}{2}\left(\ket{g}_{ii}\bra{e}+\ket{e}_{ii}\bra{g}\right)(\hat{a}+\hat{a}^{\dagger})t}\right]=0.
\end{align}
Combine the results of Eq.~\eqref{eq:b5}, we can get
\begin{align}
     P_{eg}^i-P_{ge}^i=\im Tr\left[(\ket{g}_{ii}\bra{g}\otimes\rho_{th})\rme^{\im \frac{\eta_i\Omega}{2}\left(\ket{g}_{ii}\bra{e}+\ket{e}_{ii}\bra{g}\right)(\hat{a}+\hat{a}^{\dagger})t}\right.\nonumber \\
      \left.(\im\ket{e}_{ii}\bra{g}-\im\ket{g}_{ii}\bra{e})\rme^{-\im \frac{\eta_i\Omega}{2}\left(\ket{g}_{ii}\bra{e}+\ket{e}_{ii}\bra{g}\right)(\hat{a}+\hat{a}^{\dagger})t}\right]\nonumber                 \\
     =Tr\left[\rho_{th}\frac{\im}{2}\left(\rme^{\im \frac{\eta_i\Omega}{2}(\hat{a}+\hat{a}^{\dagger})t}-\rme^{-\im \frac{\eta_i\Omega}{2}(\hat{a}+\hat{a}^{\dagger})t}\right)\right]=0.
\end{align}
Thus, $P_{eg}^i=P_{ge}^i=0$, and the reduced density matrix of the qubit only has the diagonal elements, in which case the QFI \cite{D.2011} is $F_{Q}=\frac{\partial_{\bar{n}} \rho_{11}}{\rho_{11}}+\frac{\partial_{\bar{n}} \rho_{22}}{\rho_{22}}=F_C$.

\bibliographystyle{apsrev4-1}
\bibliography{refs}

\end{document}